# COTERIES, SOCIAL CIRCLES AND HAMLETS
## CLOSE COMMUNITIES: A STUDY OF ACQUAINTANCE NETWORKS

Robert J. Mokken [1]

[1] University of Amsterdam (Faculty of Social and Behavioural Sciences: Amsterdam School of Communications Research (ASCoR), and Faculty of Science: Informatics Institute (II)).

**Abstract**

In the analysis of social networks many relatively loose and heuristic definitions of 'community' abound. In this paper the concept of closely knit communities is studied as defined by the property that every pair of its members are neighbors or has at least one common neighbor, where the neighboring relationship is based on some more or less durable and stable acquaintance or contact relation. In this paper these are studied in the form of graphs or networks of diameter two (2-clubs). Their structure can be characterized by investigating shortest spanning trees and girth leading to a typology containing just three or, in combination, six types of close communities.

## 1. Introduction

In the analysis of social structure many relatively loose and heuristic definitions of 'community' abound. Here we will focus on structural properties associated with closely-knit groups such as cliques, coteries, peer groups, primary groups and face-to-face communities, such as small villages and artist colonies. Considered as dense social networks they can serve as powerful sources of social capital and support for their members, and can serve both quick diffusion of social innovation as well as speedy contamination in epidemiology.

### 1.1. Acquaintance networks

In this paper the concept of closely knit communities is studied as defined by the property that every pair of its members are neighbors or has at least one common neighbor, where the neighboring relationship is based on some more or less stable, enduring mutual acquaintance or contact relation. Hence we shall call them 'acquaintance networks' here and think of them as a group or small community, the members of which are all acquainted with each other, or have at least one common acquaintance in that community. Such a community is closely knit because any of its members can reach or contact all other members directly through his neighbors. We shall see that, seen as a social network spanned by that relation, such close communities can be studied in the form of graphs or networks of diameter two: 2-clubs as defined by Mokken (1979). This makes it possible to discover some characteristic properties and typologies for them. As such they are an extension beyond the well-known cliques (complete graphs of diameter 1), which are associated with smaller compact groups, too small to scale at levels associated with the community notion.

In section 1.2 we shall introduce first some necessary concepts from the theory of graphs, and section 1.3 will use these to demonstrate some general properties for graphs of diameter 2, *i.e.* 2-clubs as acquaintance networks. Section 2 explains the concept of the *span* of a graph in terms of that of a smallest spanning tree (sections 2.1, 2.2 and 2.3) which will enable us to distinguish just three span-classes for acquaintance networks. Section 3 explores in more detail the properties for each of these three classes: *coteries* (section 3.1), *social circles* (section 3.2) and *hamlets* (section 3.3). Section 4 then proceeds with an additional characterization of cliquish and cliqueless acquaintance networks based on the introduction of the girth (shortest cycle) in a graph or network (section 4.1) which leads to a final span-girth typology for acquaintance networks in section 4.2. Section 4.3 considers cliqueless acquaintance networks followed by cliquish networks in section 4.4. The paper concludes in section 4.5 with the introduction of local cliquelessness.

### 1.2. Some concepts of the theory of graphs.

For the general terminology in this paper we refer to Harary (1969), while here we shall give a cursory introduction to some of the concepts to be used below.

We shall only consider simple graphs *i.e.* undirected graphs without loops or multiple edges. A simple graph consists of a set of n points $V(G)$ or $V$ and m lines $(u,v)$ connecting points $u, v \in V(G)$. $E(G)$ or $E$ denotes the set of lines of a graph, $u,v \in E(G)$ indicating that the points u and v are adjacent, *i.e.* connected by a line, in

G. The complement $\bar{G}$ of G denotes the graph with the same point set V as G, but two points u and v are adjacent in $\bar{G}$ if and only if they are not adjacent in G.

$$(u,v) \in E(\bar{G}) \text{ iff } (u,v) \notin E(G).$$

Obviously, G is the complement of $\bar{G}$.

A path joining two points u and v consists of a sequence of adjacent points and lines connecting u and v such that each point and line occurs just once. The length of a path is the number of its constituting lines. A graph is connected if every pair of its points is joined by a path. It is disconnected if there exist pairs of points not joined by a path. A graph G is totally disconnected if there are no lines in G.

The degree of a point $v_i$ in G, denoted as $d_i$, $d(v_i)$ or deg $v_i$, is the number of lines incident to that point. Any point u adjacent to v in G is called a neighbor of v. Hence in simple graphs the degree of v also denotes its number of neighbors in G.

The set of neighbors of a point u is called its neighborhood. For any pair of points u, v ∈ G we shall define

$$V_{uv}, V_{u\bar{v}}, V_{\bar{u}v} \text{ and } V_{\bar{u}\bar{v}} \quad (1)$$

as, respectively, the set of points of G (exclusive of u and v) which are:

- adjacent to both u and v, ($V_{uv}$);
- adjacent to u but not to v ($V_{u\bar{v}}$);
- adjacent to v but not to u ($V_{\bar{u}v}$);
 or
- not adjacent to either u and v ($V_{\bar{u}\bar{v}}$).

A (point) subgraph H of G consists of a point set V(H) ⊆ V(G) together with all the lines joining points u,v ∈ V(H) in G. A subgraph H of G is maximal with respect to a property if there is no larger subgraph of G with that property and containing H. For instance: a component of a graph G is a maximal connected subgraph of G. Thus a disconnected graph consists of two or more components.

The complete graph $K_n$ is the graph with n points, all pairs of points of which are adjacent. A clique of G is a maximal complete subgraph *of G*. The complete bipartite graph or bigraph $K(V_1,V_2)$ is a graph whose point set can be partitioned into two subsets $V_1$ and $V_2$ such that for all $u \in V_1$ and $v \in V_2$, the line $(u,v) \in E(G)$ and no two points in $V_i$ (i=1,2) are adjacent. The k-partite complete graph is defined accordingly.

The distance $d_G(u,v)$ between two connected points in G is given by the length of a shortest path connecting u and v in G. For each pair of points in G we have that their distance in G is at most equal to their distance in any subgraph H of G:

$$d_G(u,v) \leq d_H(u,v). \tag{2}$$

The diameter d(G) of a connected graph is given by the largest distance for a pair of points in G. Again, if H is a subgraph of G then

$$d(G) \leq d(H). \tag{3}$$

Thus the diameter of a subgraph of a graph G is at least equal to the diameter of G.

To generalize the concept of neighborhood consider all points v at distance i from u in G

$$d_G(u,v) = i; \; i = 1, 2,...,d(G).$$

then the set of points v of G for which $d_G(u,v) = i$ is called the i-neighborhood of u: $V_i(u)$. For 2-clubs, graphs of diameter 2 the 1 and 2 neighborhoods span the whole graph as $V_1(u) \cup V_2(u) = V(G) - u$.

The eccentricity of a point v in G is $\max_u d(u,v)$ in G. The radius r(G) is the minimum eccentricity of its points.

A cycle of G is a closed path in G where each point is both a starting as an endpoint in that path. Its length is the number of lines in it. The shortest cycle, a triangle, has length three.

A connected graph with no cycles (acyclic) is called a tree. An acyclic disconnected graph is called a forest.

## 1.3. Graphs of diameter two

As suggested in section 1.1 the concept of 'acquaintance network' will be associated in this paper with simple graphs of diameter at most two: $d(G) \leq 2$. Moreover, the class of cliques or complete graphs $K_n$ with diameter 1 will not be treated, or only very cursorily here. We shall therefore concentrate mainly on graphs of diameter two, the class of which shall be denoted by $D_2$.

Graphs $G \in D_2$ are easily characterised by the defining circumstance that all pairs of non-adjacent points $u, v \in V(G)$ have at least one common neighbor

$$V_1(u) \cap V_1(v) \neq \emptyset$$

A further characterisation or classification is not easy. The literature concerning graphs of diameter two ($G \in D_2$) is vast and extremely dispersed, a phenomenon which corresponds with the enormous multitude of such graphs. A few examples may suffice to make this clear. Sabidussi (1966, 584) states that if a graph $G$ of $n$ points has minimum degree $\delta(G) \geq \frac{1}{2}(n-1)$ the diameter $d(G) \leq 2$. Høivik and Gleditsch (1970; Høivik, 1969) give another result concerning a binomial random graph model where for each pair of points the probability of a line is .50. Their result implies that for very large n ("in the limit") the proportion of resulting graphs with diameter at most two ($d(G) \leq 2$)) approaches one: "almost all" graphs have diameter 2 or 1 for very large n. These examples, in themselves rather specific, point already at huge numbers and a great variety of graphs of diameter two for those cases.

A more general impression can be given by an example involving the complement $\bar{G}$ of a graph G. The complement $\bar{G}$ is the graph with the same pointset V as G, and two points u and v are adjacent in $\bar{G}$ if and only if they are not adjacent in G. Mokken (1980; Theorem 2) proves that if the diameter of $\bar{G}$ is at least four then the diameter of G itself is at most 2.

The multitude of possible graphs of diameter two is illustrated by the classification of all possible graphs on the basis of the diameter $d(G)$ and that of its complement: $\bar{d(G)}$, as given in figure 1. (Mokken, 1980, 6).

Figure 1. Connectedness and diameters of G and $\overline{G}$

|     |     | $d(\overline{G})$ |     |     |     |     |
|-----|-----|-----|-----|-----|-----|-----|
|     |     | 1 | 2 | 3 | $\geq 4$ | $\infty$ |
|     | 1   | − | − | − | − | + |
|     | 2   | − | + | + | + | + |
| (G) | 3   | − | + | + | − | − |
|     | $\geq 4$ | − | + | − | − | − |
|     | $\infty$ | + | + | − | − | − |

In figure 1 we see that the diameter condition on $\overline{G}$ that <u>least</u> restricts the diameter of G is the condition $d(\overline{G}) = 2$, as the diameter of G may then vary from two to infinity (G disconnected).

Our further investigation of acquaintance networks may be somewhat simplified by considering separable graphs.

Let G − u denote the graph consisting of the pointset V(G) − {u} and all lines in G, except those incident with u. G − u therefore is the graph resulting if we eliminate the point u and all lines of G incident with u. A connected graph G is called <u>separable</u> if there exists a point u ∈ V(G) such that G − u is disconnected. Points u of G with that property are called <u>cutpoints</u> of G. These cutpoints correspond in acquaintance networks with <u>critical members</u>, the elimination of which disrupts communication in the full network along acquaintance paths.

A connected graph is called non-separable or, for short, a <u>block</u> if it has no cutpoints. The corresponding acquaintance network has no critical members.

A graph G has a <u>spanning star</u> if there is a point v which is adjacent to all other points of G. Obviously its degree then is $d(v) = n - 1$. The point v is the center of the spanning star.

In an acquaintance network consequently a spanning star denotes a member, who is acquainted with (the neighbor of) all other members of the network. The following proposition is not difficult to establish.

<u>Theorem 1</u>. If an acquaintance network (G∈$D_2$) is separable (<u>i.e.</u> has at least one cutpoint) then G has a single spanning star with center $v_0$.

An acquaintance network therefore can have at most one critical member: a cutpoint.

Obvious corollaries are

<u>Corollary 1</u>. If G ∈ $D_2$ does not have a spanning star, then G is a block.

Non-separable acquaintance networks do not have spanning stars.
The next corollary concerns the maximum degree $\Delta(G)$ in an acquaintance network: the maximum neighborhood size $|V_1(u)|$ to be found for its members $u$.

Corollary 2. If $G \in D_2$ and $\Delta(G) < n - 1$, then G is a block.

A singleton (or pendant) of a graph is a point with degree one.
A singleton in an acquaintance network therefore has just one neighbor in that network.

Corollary 3. If $G \in D_2$ has singletons, then G has a single spanning star and all singletons are adjacent to its center $v_0$.

In acquaintance networks singletons, if present, are all neighbors of $v_0$, the center of a single spanning star in that network. They all belong to the neighborhood of the central member $v_0$. An example is given in figure 2, (b).

Figure 2. Separable acquaintance networks.

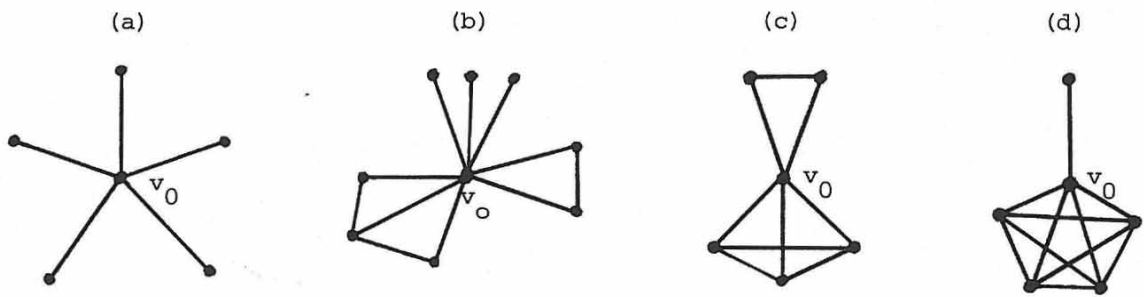

(a)  (b)  (c)  (d)

Separable acquaintance networks are therefore sufficiently characterised. They have a single spanning star, the center of which also is the only cut-point of the graph.
Case (a) gives the extreme case where G itself is a star ($S_6$) with six points and center $v_0$. Cases (c) and (d) give other extreme cases for six points, where G is separable and has the maximum number of lines possible, whithout losing separability. The center in both cases connects two complete graphs: a $K_2$ and $K_3$ in (c) and a $K_1$ and $K_4$ in case (d).

In our further investigations we shall no longer consider <u>separable</u> acquaintance networks and focus our attention on non-separable ones: <u>blocks</u>. Consequently, unless stated otherwise, we shall assume our acquaintance networks (G ∈ $D_2$) to be non-separable.

In our introductory section 1.1. we introduced acquaintance networks as compactly organized, centralized graphs and an extension beyond the well-known cliques ($K_n$ and diameter one) to graphs of diameter two. Compactness or degree of centralization of the network as a whole (Newcomb, 1961, 181), however, is only partly given by the diameter.

The great variety of graphs of diameter two, as illustrated above, necessitates a further attempt to classify acquaintance networks according to their inner global compactness. We shall suggest such a classification in terms of the concepts of <u>span</u> and <u>girth</u> of a connected graph. The resulting typology further subdivides the class $D_2$ of acquaintance networks in subclasses of varying homogeneity or compactness.

## 2. The span t(G) of G: smallest spanning trees.

We shall first recapitulate some necessary well established results concerning trees.

### 2.1. Trees

A tree is a connected graph without cycles; it necessarily has exactly n - 1 lines (m = n - 1). The addition of any other line would introduce a cycle. A point, which is a singleton is called an end-point of a tree. According to a well-known result every tree with more than two points (n ≥ 2) has at least two end-points. Each pair of points of a tree is connected by at most one path, otherwise there would be cycles.

The <u>eccentricity</u> e(u) of a point u in a connected graph is max d(u,v) for all v in G: the longest distance between u and another point of G. The <u>radius</u> r(G) is the **minimum** eccentricity of the points. Obviously, the maximum eccentricity is given by the diameter of G.

A point v is a <u>radial point</u> of G if e(v) = r(G). The <u>radial center</u> of G is the set of all radial points of G.

The radial center of a tree consists of either one, or two adjacent radial points as given in the following theorem.

Theorem 2. If a graph T is a tree then,
(1) every longest path of T contains the radial points of T;
(2) if $d(T) = 2k$ ($k=1,2,...$ ; even diameter), then its radial center is a unique radial point and

$$r(T) = \tfrac{1}{2}d(T) = k;$$

(3) if $d(T) = 2k+1$ ($k=1,2,...$; odd diameter) then its radial center consists of two adjacent radial points and

$$r(T) = \tfrac{1}{2}(d(T)+1) = k+1$$

The proof, by induction on n, is not difficult and will not be given here. It is based on the construction of the tree T', obtained by eliminating all endpoints of T and an approach to be found in Harary (1969, 35).

## 2.2. Some types of trees

We shall now introduce some specific trees, which we shall need in the sequel. We shall use in our definition the tree T' which we get from a specific tree if we eliminate all the endpoints of T.

A well-known tree is the <u>star</u> $S_n$ on n points. A star $S_n$ is any tree with n points such that $T' = K_1$, a single point. In figure 3 we give an illustration of the smallest star $S_3$ and $S_6$.

Figure 3.   Stars $S_3$ and $S_6$

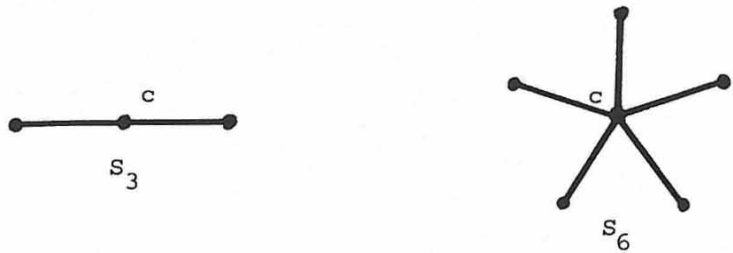

A star $S_n$ has at least 3 points ($n \geq 3$), diameter 2 and radius 1, as we can verify in figure 3. In fact it is not difficult to prove as an equivalent characterization of a star:

Theorem 3.   A tree T is a star iff its diameter $d(T)$ is two.

In conformity to theorem 2 a star has a single radial point c, which is its center.

A <u>coupled star</u> ($CS_n$) is any tree T with n points, such that $T' = K_2$, a pair of adjacent points.
In figure 4 are given some illustrations, concerning the smallest coupled star ($CS_4$) and a coupled star on nine points $CS_9$.

Figure 4.  Coupled stars $CS_4$ and $CS_9$

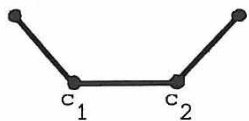
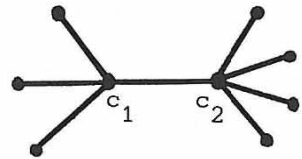

the $CS_4$          a $CS_9$

Obviously, a tree which is a coupled star should have at least four points: $n \geq 4$. Its diameter is 3, which again is a characterising condition.

Theorem 4.  A tree T is a coupled star iff its diameter $d(T) = 3$.

Again, in conformity to theorem 2, coupled stars have a radial center of two adjacent radial points $c_1$, $c_2$ (the $K_2$ of $T'$) and its radius

$$r(CS) = \tfrac{1}{2}(d(CS)+1) = 2.$$

Finally, we have to introduce the double star.
A <u>double star</u> $DS_n$ is any tree on n points such that $T' = S_k$ ($3 \leq k \leq n-2$).
Some double stars are given in figure 5. The smallest double star has 5 points and is equivalent to a single path of length 4 connecting 5 points. The other example concerns a double star on eleven points, a $DS_{11}$.

Figure 5.  Double stars $DS_5$ and a $DS_{11}$

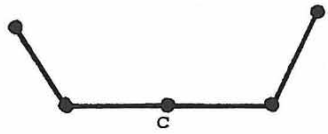

the $DS_5$

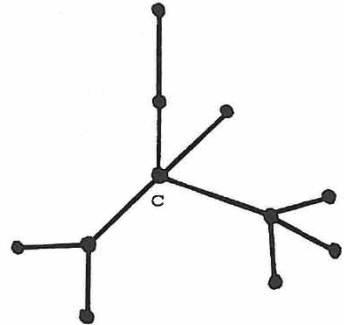

a $DS_{11}$

The diameter of a double star $DS_n$ is four and again it is not difficult to prove the equivalent characterization:

Theorem 5. A tree T is a double star iff its diameter $d(T) = 4$.

Conform theorem 2 double stars have a radius $\frac{1}{2} d(DS_n) = 2$ and a single radial point c, which, of course, is also the single central point of the star $T' = S_k$.

## 2.3. Smallest spanning trees: the span t(G) of a graph

It is well known that every connected graph G has at least one spanning tree. We may therefore consider for any connected graph G the set of its spanning trees, $T(G)$, so that

$T \in T(G)$ iff T is a spanning tree of G.

We shall say that $T(G)$ contains the trees of G. Now consider the class of graphs with given diameter

$d(G) = d$,

to be denoted as

$D_d : G \in D_d$ iff $d(G) = d$.

The diameter of G is a measure of the overall proximity of the points in G. Now, given the diameter of G, we may consider a graph G as more tightly connected than another graph G' with the same diameter if the trees of G can have smaller diameter than those of G'.

A smallest spanning tree (s.s.t.) of G is a spanning tree of G with smallest diameter taken over the trees of G. (T ∈ $T(G)$).

It is the smallest connected graph (in the sense of its diameter) which remains if one eliminates the maximum number (m − n+1) of the m lines of G, without disconnecting its n points.

Let t(G), or t, denote the minimum value of the diameter of a tree of G. We shall call t(G) the <u>span</u> of G:

$$t(G) = t = \min d(T) \text{ for } T \in T(G) \tag{4}$$

A tree of G for which d(T) = t, therefore is a smallest spanning tree (s.s.t.) or, for short, a smallest tree of G. It can be considered as a "skeleton" of G, its diameter t, the span of G, measuring the compactness of that skeleton. Which values are possible for the span t(G) of a connected graph G, given its diameter d?

That value is given by the following theorem.

<u>Theorem 6.</u>   Let a connected graph G have diameter d, (G ∈ $D_d$), then the span t(G) satisfies

$$d(G) \leq t(G) \leq 2\, d(G) \tag{5}$$

<u>Proof</u>: A smallest tree ($T_s$) of G is a spanning subgraph of G. Therefore, according to (3)

$$d(G) \leq d(T_s) = t(G),$$

which settles the lefthand inequality of (5). As by definition

$$r(G) \leq d(G) \tag{6}$$

it will be sufficient to prove t(G) ≤ 2 r(G) to establish the righthand side of (5).

We shall use an algorithm by which it is always possible to construct a tree T of G such that

$$d(T) \leq 2\, r(G)$$

Let $u_0$ be a radial point of G. Hence, for all v ∈ V(G)

$$d_G(u_0, v) \leq r$$

and there is at least one point $u_1$ ∈ V(G) such that

$$d_G(u_0, u_1) = r$$

Let $V_i(u_0)$ denote the set of points $v \in V(G)$, such that

$$d_G(u_0, v) = i;\quad i=1, 2, \ldots, r.$$

$V_i(u_0)$ is, as we saw in section 1.2., the i-neighborhood of point $u_0$ in G: the set of points at distance i of $u_0$ in G.

Obviously, none of these sets is empty.

Form the subgraph $T_1$ of G consisting of $u_0$ and all points $v \in V_1(u_0)$, cutting all lines in G which join $v_i, v_j \in V_1(u_0)$.

Enlarge this subgraph in subsequent steps in such a way, that the i-th step (i=2, ..., r) is given as follows:

form the subgraph $T_i$ of G consisting of:

(1) $T_{i-1}$;
(2) all points $v \in V_i(u_0)$;
    by
(3) cutting all lines joining $v_i, v_k \in V_i(u_0)$;
(4) cutting for each $v \in V_i(u_0)$ all lines except one joining v with points $u \in V_{i-1}(u_0)$.

This procedure ends with a spanning subgraph $T_r$ of G, by adding all the points $v \in V_r(u_0)$. From its construction it follows that

- $T_r$ is connected;
- $T_r$ has n-1 lines.

Hence $T_r$ is a spanning tree of G and its diameter is at most 2r, <u>i.e.</u> if

$$|V_r(u_0)| \geq 2.$$

Consequently, for a smallest tree $T_s$ we must have

$$d(T_s) = t(G) \leq d(T_r) \leq 2r$$

which, in view of (6), completes our proof.

### 2.4. Three span-classes for acquaintance networks

As a consequence of theorem 6, the class $D_d$ of connected graphs of diameter d, can be partitioned into subclasses $D_d(t)$, which are characterised by a specific value of t, the span of graphs belonging to that subclass.

As, given d, the span t, according to theorem 6, can assume values d, d+1, ..., 2d, there are d+1 subclasses $D_d(t)$.

For acquaintance networks, $G \in D_2$, with diameter two their span therefore allows us to partition them in three subclasses, according to the three possible values for t:

Corollary 4. The span t(G) of acquaintance networks can assume the values 2, 3 and 4.

Hence we can distinguish three subclasses of acquaintance networks, according to their span: $D_2(t)$ : t = 2, 3, 4.

## 3. Coteries, social circles and hamlets.

In terms of their span t graphs in class $D_2(2)$ are more compact than $D_2(3)$, and $D_2(4)$ is the least compact class. We shall therefore need a closer look at each of these subclasses.

### 3.1. Coteries: $G \in D_2(2)$

Acquaintance networks $G \in D_2(2)$ have a span of two. They therefore have at least one smallest tree of diameter 2, which by theorem 3 is a spanning star. Such acquaintance networks therefore contains one or more central members, with spanning stars ($S_n$) each of which is a neighbor of all other members of the network. For these reasons we might coin the name of 'coterie' for them, to distinguish them from cliques. )

We have seen, in section 1.3., that all separable graphs $G \in D_2$ belong to this subclass, having a single spanning star. We shall from now on adhere to our convention, stated in section 1.3. and consider, unless otherwise stated, only blocks, non-separable acquaintance networks.

Blocks belonging to the subclass of acquaintance networks with span t=2 may range from the cases of networks with a single spanning star to the case where G is almost a clique.

Figure 6. Examples of blocks with span t=2.

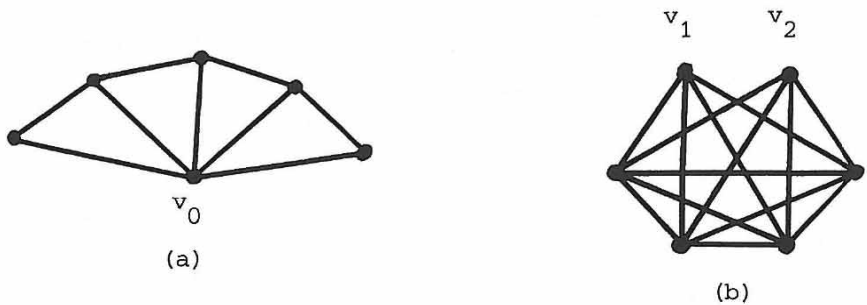

(a)   (b)

They are illustrated in figure 6, where (a) gives a minimal case: on elimination of the central point of the single spanning star a tree remains. The maximal case is illustrated in (b) of figure 6: if the line $(v_1, v_2)$ had been present we would have a clique. $(K_6 - (v_1, v_2))$.

It will be clear, that if an acquaintance network is not a <u>coterie</u>, then it will not have a spanning star. For every point of G the degree will be less than n-1. In other words, its maximum degree $\Delta(G) < n-1$. The acquaintance network then will be either a <u>social circle</u> ($G \in D_2(3)$) or a hamlet ($G \in D_2(4)$).

<u>Corollary 5.</u>  Iff $G \in D_2$, then $G \in D_2(3)$ or $G \in D_2(4)$ iff $\Delta(G) < n-1$.

Social circles and hamlets are acquaintance networks characterized by the fact that they have no members which are a neighbor to all other members (points) of the network.

Moreover, by corollaries 2 and 5, all social circles and hamlets are non-separable acquaintance networks.

<u>Corollary 6.</u>  If $G \in D_2(3)$ (social circles) or $G \in D_2(4)$ (hamlets) then G is a block.

Graphs of diameter 2 with span t equal to 3 or 4 are blocks.

We may also note that <u>singletons</u> do not occur in social circles and hamlets, as a consequence of corollary 3. Therefore, for all social circles or hamlets the minimum degree is at least two

$$\delta(G) \geq 2; \tag{7}$$

all their members have at least two neighbors or acquaintances.

A combination of (7) and corollary 5 gives for the degrees of all points u of social circles ($G \in D_2(3)$) and hamlets ($G \in D_2(4)$)

$$2 \leq d(u) \leq n-1 \tag{8}$$

3.2. Social circles: $G \in D_2(3)$.

An acquaintance network $G \in D_2(3)$ has span $t(G) = 3$ and consequently at least one smallest tree with diameter 3. By theorem 4 that s.s.t. is a coupled star $CS_n$. (See figure 4).

Consequently, that coupled star has a radial center consisting of two adjacent points u and v of G. From the construction of the s.s.t. $CS_n$ as indicated in the proof of theorem 6 and as illustrated by figure 4, these two adjacent members u and v are together neighbors (or acquaintances) of all the other points (members) of the acquaintance network.

We may designate such a radial central pair of points on a s.s.t. of G as a <u>pair of central neighbors</u> of G.

We may tentatively call acquaintance networks with span t=3 <u>social circles</u>, borrowing the term from Kadushin, who used the concept for more general networks. (Kadushin, 1974; Alba and Kadushin, 1976).

<u>Social circles do not have spanning stars, but they have pairs of central neighbors</u>. We shall formulate some of their properties below.

Let, as in (1) of section 1.2., for any pair of points u, v of G, $V_{\bar{u}v}$ denote the set of points which are adjacent to v in G, but not to u and let $V_1(u)$ denote the neighborhood of u. Then, from the fact that a social circle ($G \in D_2(3)$) does not have a spanning star and from corollary 5 we immediately establish:

<u>Property 1</u>.  If $G \in D_2(3)$ then for all points $u \in V(G)$ there is a point $v \in V_1(u)$ such that

$$V_{\bar{u}v} \neq \emptyset \qquad (9)$$

Each member u of a social circle has at least one neighbor, with a neighbor which is not an acquaintance of u.

Expression (9) immediately implies that for all points $u \in V(G)$

$$V_2(u) \neq \emptyset \qquad (10)$$

For all points u of G there are points at distance two in v.

Let according to (1) of section 1.2. $V_{\bar{u}\bar{v}}$ denote the set of points which are adjacent neither to u nor to v in G.

<u>Property 2</u>.  Let the adjacent pair of points u and v be a pair of central neighbors of $G \in D_2(3)$ then

$$V_{\bar{u}\bar{v}} = \emptyset \qquad (11)$$

Expression (11) formalizes our remark above that u and v together are neighbors to all the other members of the acquaintance network i.e.

$$V(G) = V_1(u) \cup V_1(v) \qquad (12)$$

An acquaintance network is a social circle if it has no spanning stars and there is at least one member (point) u with at least one neighbor v, such that there are no other members in G which are not neighbors of u or v.

### 3.3. Hamlets: $G \in D_2(4)$

An acquaintance network $G \in D_2(4)$ has span $t(G) = 4$ and consequently at least one s.s.t. with diameter 4. By theorem 5 such a tree is a double star $(DS_n)$ as illustrated in figure 5.

The reader can easily verify, with the help of the construction referred to in the proof of theorem 6, that for each point of $G \in D_2(4)$ a $DS_n$ can be found with that point as a center.

$G \in D_2(4)$ has no spanning stars, nor has it a pair of central neighbors. The proximity of its points therefore is on the whole less than in coteries and social circles and is reminiscent of a small village community. I shall therefore refer to acquaintance networks of span 4 ($G \in D_2(4)$) as <u>hamlets</u>. As hamlets have no spanning stars property 1 and the expressions (9) and (10) are also valid for them.

As they also do not have a pair of central neighbors the following property is characteristic for hamlets.

<u>Property 3.</u>   Let $G \in D_2(4)$ then for every pair of adjacent points $u, v \in V(G)$

$$V_{\bar{u}\bar{v}} \neq \emptyset \tag{13}$$

For each pair of neighbors (adjacent) u and v of a hamlet G there is always at least one member of G which is a neighbor neither of u nor of v.

We can conclude our discussion with a stronger characterisation theorem for acquaintance networks.

<u>Theorem 7.</u>   Let $G \in D_2$, then $G \in D_2(3)$ (social circle) or $G \in D_2(4)$ (hamlet) iff for all points $u \in V(G)$ there is a point $v \in V_1(u)$ such that

$V_{\bar{u}v} \neq \emptyset$;

and

(1) $G \in D_2(3)$ (social circle) iff there is a point $u_0$ and a point $v_0 \in V_1(u_0)$ such that

$V_{\bar{u}_0 \bar{v}_0} = \emptyset$; i.e. $V(G) = V_1(u_0) \cup V_1(v_0)$;

(2) $G \in D_2(4)$ iff for all points $u \in V(G)$ and for all $v \in V_1(u)$

$V_{\bar{u}\bar{v}} \neq \emptyset$.

The proof can be based for the first part on the fact that if G ∈ $D_2$ then G ∈ $D_2(2)$ iff Δ(G) = n-1 and for (1) and (2) on the propositions established thus far.

## 4. Cliquish and cliqueless acquaintance networks: span-girth typology

Our investigation of acquaintance networks (2-clubs, Mokken, 1979) regarding their compactness, or the overall proximity of their members has thus far been in terms of their skeleton: smallest spanning trees as measured by their diameter, the span t(G).
We could add to that study an analysis of their inner, circular communication patterns as a related but different indication of that proximity or compactness. We shall do so by looking at their cycles, more specifically shortest cycles and their <u>girth</u>, the length of a shortest cycle. This will enable us to extend our typology of acquaintance networks to one based on both the span and the girth of graphs of diameter 2.

### 4.1. Smallest cycles: the girth of a graph.

The girth g(G) of a connected graph G is given by the length of a shortest cycle in G (Harary, 1969, 13). Note that the girth is defined for connected graphs and not for trees, as these are acyclic graphs. Furthermore, in simple graphs cycles can have only have length greater than 2 so that
$$g(G) \geq 3.$$
Hence the minimum value of g(G) is three. The maximum value of the girth depends on the span of the graph G, as given by the following theorem.

<u>Theorem 8</u>. Let G be a connected (cyclic) graph with diameter d and let t(G) be its span, then
$$3 \leq g(G) \leq t(G) + 1 \leq 2d(G) + 1.$$
<u>Proof</u>. Let $v_g$ denote a shortest cycle of length g = g(G) in G. Then in every tree T of G this cycle is induced through the addition of one line of G to T. The elimination of this line leaves a unique path of length g-1 in T. Hence for every T of G we have g-1 ≤ d(T).

As this includes smallest trees of G we have

$$g-1 \leq t(G)$$

From theorem 6 we have

$$t(G) \leq 2 d(G),$$

so that

$$g \leq t(G) + 1 \leq 2 d(G) + 1,$$

which completes our proof.

As a consequence of theorem 8 we can for a class of (not acyclic) graphs of given finite diameter d, $G \in D_d$, distinguish subclasses characterized by a given span $t(G) = t$ and girth $g(G) = g$, to be denoted by

$$G \in D_d(t,g).$$

These classes form a typology of the graphs of given diameter: their <u>span-girth typology</u>.

According to theorem 6:

$$d \leq t(G) \leq 2d$$

and according to theorem 8:

$$3 \leq g(G) \leq t(G) + 1.$$

Hence for given $t(G) = t$ there are $t-1$ subclasses, so that the partitioning of the class $D_d$ of graphs of diameter d involves

$$\sum_{t=d}^{2d} (t-1) = \tfrac{1}{2} (d+1)(3d-2) \qquad (14)$$

subclasses.

## 4.2. Span-girth typology of acquaintance networks

It is well known that the only acyclic graphs (trees) of diameter two are the stars $S_n \in D_2(2)$. From theorem 8 it follows that acquaintance networks, $G \in D_2$, unless they are stars $S_n$, can have girth 3, 4 or 5. Their shortest cycles therefore are the triangle $C_3$, the building stone of cliques; the square $C_4$, a cycle on four points; and the pentagon $C_5$, the cycle on five points.

More precisely, we have according to their span and girth:

The typology of figure 7 now can be presented in an adapted version as in figure 8.

Figure 8.  Cliquish and cliqueless acquaintance networks

|  | span | Girth | |
|---|---|---|---|
|  |  | g = 3 Cliquish | g = 4 Cliqueless |
| Coteries | t = 2 | $D_2(2,3)$ | – |
| Social circles | 3 | $D_2(3,3)$ | $D_2(3,4)$ |
| Hamlets | 4 | $D_2(4,3)$ | $D_2(4,4)$ |

What do the various types of cliquish and cliqueless acquaintance networks look like?

4.3. Cliqueless acquaintance networks

Except for the separable single case of stars $S_n$, there are no <u>cliqueless coteries</u>. All coteries are cliquish.

There can be <u>cliqueless social circles</u>, $G \in D_2(3,4)$, but they can be shown to be of a very special type: they coincide with the complete bipartite graphs.
A complete bipartite graph G is a graph whose point set V can be partitioned into two subsets $V_1$ and $V_2$ such that
every point of $V_1$ is joined by a line of G with every point of $V_2$ and there are no lines in G joining pairs of points of $V_1$ or of $V_2$.
We shall denote a bipartite graph as $K(V_1, V_2)$. Alternatively, if $V_1$ has k points en de $V_2$ n-k points, we may write $K(k_1, n-k)$; $(k_1, n-k \geq 4)$.
Obviously, the number m of lines of G then is $k(n-k)$.
In figure 9 we give an example of a complete bipartite graph with K(2,3).

Figure 9.  Complete bipartite graph K(2,3).

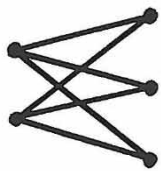

- coteries, $G \in D_2(2)$: $g(G) = 3$ (excluding $S_n$);
- social circles, $G \in D_2(3)$: $g(G) = 3$ or 4;
- hamlets, $G \in D_2(4)$: $g(G) = 3, 4,$ or 5.

The span-girth subclasses $D_2(t, g)$ therefore can be illustrated in figure 7.

Figure 7.   Span-girth typology $D_2(t, g)$.

|  | span | Girth 3 | 4 | 5 |
|---|---|---|---|---|
| coteries: | t = 2 | $D_2(2,3)$ | - | - |
| social circles | 3 | $D_2(3,3)$ | $D_2(3,4)$ | - |
| hamlets | 4 | $D_2(4,3)$ | $D_2(4,4)$ | $D_2(4,5)$ |

The empty cells in figure 7 indicate alternatives which are impossible according to theorem 8.

Moreover, the class $D_2(4,5)$, hamlets with shortest cycles $C_5$, i.e. the narrowest circular communication involving five members, seem to be uninteresting for social analysis as well.

Singleton (1968) proved that if for any graph G with diameter d, its girth $g(G) = 2d + 1$, then that graph is regular. A graph is called regular if all its points have the same degree. The regular graphs in $D_2(4,5)$ correspond to Moore-graphs of diameter 2. Hoffman and Singleton (1960) have shown that there are just three of such graphs: the pentagon $C_5$, the 10 point Petersen graph of degree 3 (Harary, 1969, 89), and one graph with 50 points and degree 7. There may be still one other such graph (with 3 250 points and degree 57) but its existence has not yet been decided.

For these reasons the subclass $D_2(4,5)$ is not very relevant for our analysis of acquaintance networks so that we shall not consider them further.

Acquaintance networks with girth 4 do not have triangles, as their shortest cycles are squares $C_4$. As the triangle $C_3$ is the smallest proper clique, we shall call these acquaintance networks cliqueless.

The other types with girth 3 do have at least one triangle. Usually these triangles will be part of larger cliques. In that sense we referred to triangles as the building stones of cliques. Therefore we shall call acquaintance networks with girth 3 cliquish.

According to a well-known theorem of König, all cycles of bipartite graphs are even. (Harary, 1969, 18).

Evidently, all pairs of mixed points u, v of V, u ∈ $V_1$ and v ∈ $V_2$ are pairs of central neighbors.

Cliqueless hamlets, G ∈ $D_2(4,4)$ are not easy to characterize and are very numerous. We can attempt here only to give a number of properties.

All cliqueless hamlets have at least one cycle $C_5$. Consequently hamlets consist of at least 6 points: n ≥ 6. For an example see figure 10.

Figure 10.   Cliqueless hamlet of six points.

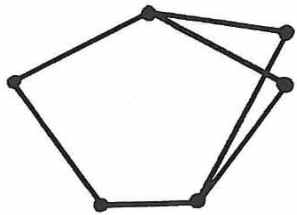

Moreover, the following properties are evident for cliqueless hamlets (G ∈ $D_2(4,4)$): for each adjacent pair of points, i.e. for each pair of neighbors:

(a) $V_{u\bar{v}} \neq \emptyset$ and $V_{\bar{u}v} \neq \emptyset$; (otherwise there would be singletons in G);

(b) $V_{u,v} = \emptyset$; (otherwise there would be triangles in G);

(c) $V_{\bar{u}\bar{v}} \neq \emptyset$; (otherwise t(G) = 3 and u, v would be a pair of central neighbors of G).

As a consequence of (c):

(d) each adjacent pair of points and therefore each line of G is on a cycle $C_5$.

A diagonal of a cycle is a line joining two non-adjacent points on that cycle. An open cycle is a cycle without diagonals. As there are no triangles:

(e) All cycles $C_4$ and $C_5$ are open.

Cliqueless hamlets therefore can be looked at as mixtures of open cycles $C_4$ and $C_5$.

(f) For each point u ∈ V(G) of a cliqueless hamlet the subgraph on $V_1(u)$ is totally disconnected; (otherwise there would be triangles in the neighborhood of u, involving u).

It seems not to be possible to characterize cliqueless hamlets more clearly.
This has thus far only been possible for specific cases, mainly in the
area of extremal graphtheory. For instance, Harary, Mokken and Tindell
identified certain types as minimal blocks of diameter two, of which the
graph of figure 10 is an example.

Cliqueless social circles and hamlets, although theoretically abundant, do
not seem empirically relevant in social investigation, except perhaps in
the area of social design and purposive organization.
In the actual empirical investigation of social structure we probably should
face the fact, that cliquishness is virtually a universal phenomenon.

4.4. Cliquish acquaintance networks.

We have seen above, that except for the single case of the star $S_n$, which
is acyclic, all coteries are cliquish:

$G \in D_2(2) \subsetneq G \in D_2(2,3)$ or $G$ is $S_n$.

We have seen in section 3.1. that they all have spanning stars.

Cliquish social circles, $G \in D_2(3,3)$ do not have spanning stars but they
have pairs of central neighbors and at least one triangle, $C_3 = K_3$, the
complete graph on 3 points.
A complete k-partite graph G is a graph whose pointset V can be partitioned
into k subsets $V_1$, $V_2$, ..., $V_k$ such that for all i, j (i≠j; i,j = 1,2,...,k)
every point of $V_i$ is joined by a line of G with every point of $V_j$ and
there are no lines in G joining pairs of points of $V_i$ or of $V_j$.
For all i (i = 1,2,...,k) the subgraphs in G on the pointsets $V_i$ are
totally disconnected. Note that the complete bipartite graph $K(V_1,V_2)$ of
section 4.3. is a special case of a complete k-partite graph.
The reader may readily verify that for k ≥ 3 all complete k-partite graphs
are cliquish social circles.
For each value of i and j (i≠j) a pair of points $u \in V_i$, $v \in V_j$ is a pair
of central neighbors.
Moreover, all acquaintance networks ($G \in D_2$) which do not have spanning
stars, but contain complete k-partite graphs (k ≥ 2) as proper spanning
subgraphs, also are cliquish social circles.
All these examples do not exhaust all possibilities of cliquish social
circles, the class of which is much larger and difficult to characterize
any further.

Cliquish hamlets (G ∈ $D_2(4,3)$) do not have spanning stars nor pairs of central neighbors. Obviously they have at least one triangle. As given in (2) of theorem 7 for all points u ∈ V(G) and for all v ∈ $V_1(u)$ we have $V_{\overline{uv}} \neq \emptyset$, which is characteristic for hamlets in general. In other words: for each pair of neighbors u, v, there is at least one point in G which is neither a neighbor of u nor of v. As a consequence each pair of neighbors, or line of G, is on a pentagon $C_5$. Evidently the cycles $C_4$ and $C_5$ need not be open and can have diagonals.
Cliquish hamlets are difficult to characterize further than that.

## 4.5. Local cliquelessness.

There is another property which distinguishes cliquish hamlets ($D_2(4,3)$) from cliquish social circles ($D_2(3,3)$). It is a property which may be denoted as local cliquelessness.

Figure 11.   Cliquish hamlets with and without local cliquelessness

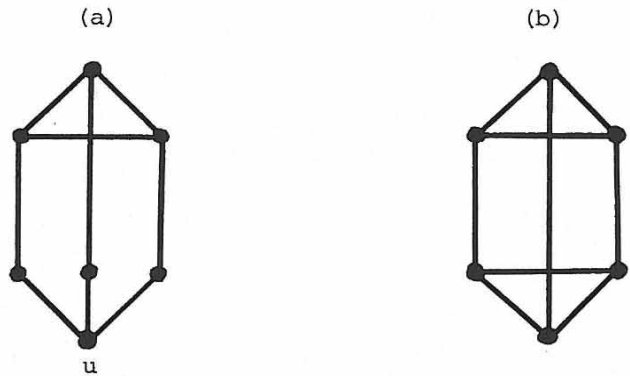

(a)          (b)

u

A point or member u of an acquaintance network G ∈ $D_2$ is said to be cliqueless if it is not a member of any clique of G. Hence a cliqueless point u is not a point on any triangle $C_3$ of G. The subgraph in G on the neighborhood of u, the $V_1(u)$ then is totally disconnected. The subgraph on u together with $V_1(u)$ is a star.

A point or member u of an acquaintance network G is said to have a cliqueless neighborhood if all neighbors of u in G are cliqueless. It will be evident to the reader that if u has a cliqueless neighborhood, then u itself is cliqueless. For cliqueless acquaintance networks (G ∈ $D_2(3,4)$, $D_2(4,4)$ or $D_2(4,5)$) obviously all points have cliqueless neighborhoods.

An acquaintance graph G is said to have the property of local cliquelessness if G has a point u with a cliqueless neighborhood.
It can be proven that this property, apart from the cliqueless acquaintance networks, is specific for cliquish hamlets.

Theorem 9. Let G be a cliquish acquaintance network and let there be a point u in G with a cliqueless neighborhood in G; than G is a cliquish hamlet: $G \in D_2(4,3)$.

The proof, though not difficult, will not be given here for reasons of space. An example of a cliquish hamlet with seven points and local cliquelessness is given in figure 11 (a).
An obvious corollary is

Corollary 7. If an acquaintance network is a cliquish social circle ($G \in D_2(3,3)$) or a coterie ($G \in D_2(2,3)$ except for $S_n$) then there are no points or members of G with cliqueless neighborhoods.

Hence in cliquish social circles ($G \in D_2(3,3)$) in particular all points u have cliquish neighborhoods.
That there can be cliquish hamlets without the property of local cliquelessness is illustrated in figure 11 (b).